\documentclass[a4paper,12pt]{revtex4}
\pdfoutput=1
\usepackage{epsfig}
\usepackage{amssymb}
\usepackage{amsfonts}
\usepackage{amsmath}
\usepackage{euscript}
\usepackage{verbatim}
\usepackage{latexsym}
\usepackage{graphicx}
\usepackage{caption}
\usepackage{float}
\usepackage{xcolor}
\usepackage{subcaption}
\usepackage{ulem}
\usepackage[colorlinks=true]{hyperref}
\usepackage{cleveref}

\newif\ifdtup

\catcode`\@=11

\@addtoreset{equation}{section}

\def\@normalsize{\@setsize\normalsize{15pt}\xiipt\@xiipt
\abovedisplayskip 14pt plus3pt minus3pt%
\belowdisplayskip \abovedisplayskip
\abovedisplayshortskip \z@ plus3pt%
\belowdisplayshortskip 7pt plus3.5pt minus0pt}

\def\small{\@setsize\small{13.6pt}\xipt\@xipt
\abovedisplayskip 13pt plus3pt minus3pt%
\belowdisplayskip \abovedisplayskip
\abovedisplayshortskip \z@ plus3pt%
\belowdisplayshortskip 7pt plus3.5pt minus0pt
\def\@listi{\parsep 4.5pt plus 2pt minus 1pt
     \itemsep \parsep
     \topsep 9pt plus 3pt minus 3pt}}

\relax

\catcode`@=12

\catcode`\@=11

\def\section{\@startsection{section}{1}{\z@}{3.5ex plus 1ex minus
   .2ex}{2.3ex plus .2ex}{\large\bf}}

\def\SymBoxes#1#2#3#4{\newdimen\un@t \un@t#3%
\raisebox{#1}{\rule{#2\un@t}{#4}\hskip-#2\un@t
\@tempdimb\un@t \advance\@tempdimb by-#4\@tempcntb#2\relax%
\@whilenum{\@tempcntb>0}\do{
\rule{#4}{\un@t}\hskip\@tempdimb \advance\@tempcntb by\m@ne}%
\hskip-#2\un@t \rule[\un@t]{#2\un@t}{#4}%
\rule[\un@t]{#4}{#4}\hskip-#4
\rule{#4}{\un@t}}\hskip-#4}                

\begin{document}

\newcommand{\beq}{\begin{equation}}
\newcommand{\eeq}{\end{equation}}
\newcommand{\bea}{\begin{eqnarray}}
\newcommand{\eea}{\end{eqnarray}}
\newcommand{\beas}{\begin{eqnarray*}}
\newcommand{\eeas}{\end{eqnarray*}}
\newcommand{\defi}{\stackrel{\rm def}{=}}
\newcommand{\non}{\nonumber}
\newcommand{\bquo}{\begin{quote}}
\newcommand{\enqu}{\end{quote}}
\renewcommand{\(}{\begin{equation}}
\renewcommand{\)}{\end{equation}}
\def \eqn#1#2{\begin{equation}#2\label{#1}\end{equation}}

\newcommand{\blue}[1]{\textcolor{purple}{#1}}
\newcommand{\red}[1]{\textcolor{red}{#1}}
\newcommand{\ket}[1]{\left| #1 \right\rangle}
\newcommand{\bra}[1]{\left\langle #1 \right|}
\newcommand{\hil}{\mathcal{H}}
\newcommand{\tr}[1]{\text{Tr}\left[#1\right]}
\newcommand{\ptr}[2]{\text{Tr}_{#1}\left[#2\right]}
\newcommand{\povms}{POVMs~}
\newcommand{\sys}[1]{$ \mathcal{#1} $ }
\newcommand{\op}[1]{\mathcal{#1} }
\newcommand{\e}[1]{\text{e}^{#1}}
\newcommand{\ee}[1]{\exp\left({#1}\right)}
\newcommand{\xb}{\overline{x}}
\newcommand{\pb}{\overline{p}}

	
%
	\title{Imperfect Measurements and Conjugate Observables}
	\author{Adarsh S} 
	\email{s.adarsh4812@gmail.com}
	\author{P. N. Bala Subramanian} \email{pnbala@nitc.ac.in}
	\author{T. P. Sreeraj}  \email{sreerajtp@nit.ac.in}
	\affiliation{Department of Physics, National Institute of Technology Calicut, Kozhikode, India.} 
%
%
%
%
%
	\renewcommand{\thefootnote}{\arabic{footnote}}

		
%
%
%

\begin{abstract}
	In the standard von Neumann interaction used in Quantum measurements, the chosen observable to which the environment (apparatus) entangles is exactly reproduced in the state of the environment, thereby decohering the quantum system in the eigenbasis of the observable. We relax this by allowing for imperfect {\it measurements} whereby the environment evolves to a state that approximately, but not exactly, reflects the state of the system. 
	In this scheme it is possible to attain approximate decoherence of conjugate quantities that resembles classical physics, which we demonstrate using an example.
	
\end{abstract}
	\vspace{1.6 cm}
	\vfill
\maketitle

\setcounter{footnote}{0}
\section{Introduction}

Our general intuition of the world around us is built into a paradigm called classical physics. Although fundamentally quantum mechanical, these systems exhibit generic behavior which we usually call as classical, and can be thought of as consisting of at least the following features:

\begin{enumerate}
	\item A perfect measurement of any observable of the system does not interfere with the measurement of any other. In particular, a measurement of generalized coordinates do not interfere with that of  conjugate momenta. Since any other observable maybe written in terms of them, they may be used to label the state of the system. Equivalently, state of the system maybe denoted by a point in the phase space. This necessarily implies that there are no entangled states.
	\item State of the system obeys Hamilton's equations of motion and given the state of system at any instant of time, it follows a unique trajectory in phase space called the classical trajectory. 
	\item State of the system cannot be in a linear combination of eigenstates of observables. In particular, it cannot be a linear combination of eigenstates of generalized coordinates or conjugate momenta.
\end{enumerate}
The above properties are only expected to emerge in a 'statistical' sense which then leads to classical mechanics when one only has access to 'coarse grained'  measurements made at energy (length) scales, which are in some sense small (large). 
The question of whether all these features imply each other or at least emerge at the same energy scales for all quantum systems is not completely understood and requires investigation. 

Quantum decoherence \cite{schlosbook,schlos,TS,zeh,zurek physt,zurekrmp2003,joosbook,Bacc,zurekprd1,zurek1,zurekprl} posits that real quantum systems interacts with environment and tend to get entangled with the environment.  It further says that if one can only measure local properties of the system, this entanglement information is essentially lost and one is left with a statistical description. With the loss of this information, system state becomes a mixed state described by a reduced density matrix.  All the statistical information about local observables of the system are encoded in this reduced density matrix. This process can be envisioned as a continuous measurement of a system observable, say ${\mathcal O}$, by the environment where an environment observable mimics ${\mathcal O}$ and acts as what we call a {\it read-off} variable. Under a clearly resolved precise measurement, eigenstates of the read-off variable associated with eigenstates of ${\mathcal O}$ become mutually orthogonal. This in turn leads to the vanishing of off diagonal elements of reduced density matrix written in the observable eigenbasis. Such a reduced density matrix maybe thought of as representing an statistical ensemble where each copy of the system is in an eigenstate of the observable and not in a linear combination of them. In a general statistical ensemble, the probability of getting a particular value $o_i$ of an observable ${\mathcal O}$, $P(o_i)=\sum\limits_i P(|\psi_i\rangle) P(o_i/|\psi_i\rangle)$, where $P(|\psi_i\rangle)$ is the probability of picking a state $|\psi_i\rangle$ from the ensemble (which corresponds to the statistical contribution to the probability) and $P(o_i/|\psi_i\rangle)$ is the conditional probability of getting $o_i$ given one has picked a state $|\psi_i\rangle$ (which is a fundamental probability coming from quantum mechanics). A diagonal reduced density matrix therefore implies that the conditional probability  is one leading to the loss of all quantum interference effects of the observable. If the interaction between system and the environment is such that the observable monitored by the environment is the generalized coordinate $X$ of the system, $X$ decoheres while the corresponding generalized momentum doesn't.

However, such exact $X$ measurements requires localization of a probe to infinitely small distances or infinitely high energies. Consequently, all real measurements are imprecise measurements. In this paper, we study the possibility of approximate decoherence of conjugate quantities under imprecise measurements by the environment which is modeled using a modified von Neumann interaction Hamiltonian.  In section (\ref{s:rev}), we review quantum mechanical measurement formalism for precise measurements. We then generalize it to smeared von Neumann measurements in \ref{s:von} and study smeared measurements for the example of a Gaussian wavefunction in \ref{s:eg}. We then analyze various domains and conclude that there are domains where $X$ and $P$ approximately decoheres thereby effectively leading to classicality in the sense of the criterion 3 above.    

\section{A short review of Quantum measurements }
\label{s:rev}
Based on the postulates of Quantum Mechanics, we can define the quantum measurement of an observable \cite{VNold,wheelerbook,adler}  by a set of (measurement) operators $ \{M_k\}_1^N $ acting on the system, with $ \sum_k M_k^\dagger M_k = \mathbf{1} $. A general state in the Hilbert space, $ \ket{\psi} \in \hil $, instantaneously after the measurement will become \[ \ket{\psi} \rightarrow \dfrac{M_k\ket{\psi}}{\sqrt{p_k}}, \ \ \text{where  } p_k = \bra{\psi}M_k^\dagger M_k\ket{\psi} \geq 0.\]
 In order to treat fundamental Quantum mechanical probabilities and statistical probabilities,  coming from incomplete knowledge of the state of the system, in the same footing, one could use density operators to specify the state of the system. In the language of density operators , the action of the measurement operator is \[ \rho \rightarrow \dfrac{M_k \rho M_k^\dagger}{p_k} , \ \ \text{where  } p_k = \tr{\rho M_k^\dagger M_k} \]
It can be easily seen that  projective measurement  is a subset of generalized measurements in which the operators $\{M_k=\mathbb{P}_k\}$ are Hermitian, i.e. $ \mathbb{P}_k^\dagger = \mathbb{P}_k $ and satisfy $ \mathbb{P}_k \mathbb{P}_l = \delta_{kl} \mathbb{P}_k $. This gives
\begin{align}
	\mathbb{P}_k^\dagger \mathbb{P}_k = (\mathbb{P}_k)^{2} = \mathbb{P}_k, \ \ \text{and~} p_k = \bra{\psi}\mathbb{P}_k \ket{\psi}.
\end{align}

Using the generalized measurements (not necessarily projective), we can define operators $ \{E_i\} $ that are Hermitian and positive semidefinite known as Positive Operator Valued Measures, or \povms in short. That is, the \povms are $ E_i = M_i^\dagger M_i $, which satisfy $ E_i^\dagger = (M_i^\dagger M_i)^\dagger = E_i $ and $ \sum_i E_i = \mathbf{1} $. Furthermore, 
\begin{align}
	\bra{\psi} E_i \ket{\psi}=  \bra{\psi} M_i^\dagger M_i \ket{\psi} = || M_i\ket{\psi} ||^2 \geq 0, \  \forall \ket{\psi } \in \hil
\end{align}
Corresponding to each eigenvalue $\lambda_i$ of an observable, there are  Hermitian operators $ \{E_i\} $, with the associated probability of measurement of $\lambda_i$ given by $ p_i = \tr{E_i \rho} $.

Even though generalized measurement is the more general notion of measurement, it can be shown that \cite{barnett,peres} any generalized measurement can be realized as a projective measurement on the system $\cal{S}$ coupled with another system $\cal {A}$ which is called an ancilla. Let's denote such a combined system $\cal{SA}$. Consider a separable state of the $\cal{SA}$, $\rho =  |\psi\rangle |\alpha\rangle \langle \alpha| \langle \psi|$. Under unitary evolution, 
\begin{align}
|\psi\rangle |\alpha\rangle \langle \alpha| \langle \psi| \rightarrow U|\psi\rangle |\alpha\rangle \langle \alpha| \langle\psi|  U^\dagger
\end{align}
Consider a projective measurement of observables $ \op{O} \otimes \op{O}_A$ of $\cal{SA}$. Let  $|\lambda_i\rangle$ and $|\alpha_j\rangle$ are eigenstates of $\op{O}$ and $\op{O}_A$ with eigenvalues $\lambda_i$ and $\alpha_j$ , then the probability of getting eigenvalues $\lambda_i$ on measurement of $ \op{O}$ is given \cite{barnett} by

\begin{align}
 P(\lambda_i)= \sum\limits_j P(\lambda_i,\alpha_j)= \sum\limits_j Tr_S(E_{\lambda_i \alpha_j }\rho') = Tr_S(E_{\lambda_i} \rho')
\end{align}
where $\rho'=|\psi\rangle \langle \psi|, E_{\lambda_i \alpha_j }= \langle \alpha |U^\dagger | \lambda_i \alpha_j \rangle\langle \lambda_i \alpha_j|U|\alpha\rangle$ and $E_{\lambda_i}= \sum\limits_j \langle \alpha |U^\dagger | \lambda_i \alpha_j \rangle \langle\lambda_i \alpha_j|U|\alpha\rangle$ and $Tr_S$ denotes trace over system states alone. Since, $E_{\lambda_i}$ is hermitian and $\sum\limits_i E_{\lambda_i}=1$, $E_{\lambda_i}$ is a POVM.

\subsubsection*{Read-off states and von Neumann Measurement}

The process of measurement of a quantum system should ideally be described by a fully quantum mechanical treatment of the system as well as the apparatus. Suppose we have a system $ \cal{S} $, of which we would like to measure an observable associated to the operator $ \cal{O} $, we may do so by use of a measuring apparatus $ \cal{A} $ with the following property: let the operator $ \cal{O} $ have a spectrum (non-degenerate for simplicity) with eigensystem $ \{\lambda_{i},\ket{\lambda_{i}}\} $, then the apparatus $ \cal{A} $ has suitable orthonormal states $ \ket{\alpha_i} $ which take eigenvalues $ a_i $ that are in one-to-one correspondence with the eigenvalues $ \lambda_{i} $ of $ \cal{O} $. We call such states \emph{read-off} states. The inherent assumption is that different $ \ket{\alpha_i} $ states are macroscopically distinguishable. It can thus be seen that, without any loss of generality, one can consider any measurement of an observable $\op{O}$ as a von Neumann measurement, consisting of (i) an entangling step which takes the apparatus to a read-off state and (ii) a reduction step where the measurement is taken.

With this prescription, consider the combined system and apparatus $ \cal{SA} $ as a closed system, which is described on the Hilbert space $ \hil_S\otimes \hil_A $, and unitarily evolves by the Hamiltonian 
\begin{align}
H = H_{\cal S}+ H_{\cal A} + H_{int}.
\end{align}
At time $ t=0 $ when the interaction between the system and apparatus are turned on, let the system be in a state $ \ket{\psi} = \sum_i c_i \ket{\lambda_{i}} $, and the apparatus in $ \ket{\alpha} $. For simplicity, we further assume that the system and apparatus self-Hamiltonians are zero (in other words the interaction Hamiltonian $ H_{int} $ is very dominant). For the apparatus to realistically reflect the state of the system, the unitary evolution should maximally entangle the system and apparatus whereby the apparatus state evolves to a read-off state, which can be accomplished by an interaction Hamiltonian of the von Neumann form
\begin{align}
	H_{int} = \sum_{i} \ket{\lambda_{i}}\bra{\lambda_{i}} \otimes A_i \ \ ,
\end{align}
where $ A_i $ is an operator that evolves the apparatus state to a read-off state in time $ T $, i.e.
\begin{align}
	\e{-i\, A_i T}\ket{\alpha} = \ket{\alpha_i}.
\end{align}
Thus, the system and apparatus together will evolve as follows
\begin{align}
	\rho(T) &= \e{-i\, H_{int}T} \rho(0)  \e{i\, H_{int}T} = \sum_{j,k} \e{-i\, H_{int}T} c_j c_k^*\ket{\lambda_{j}}\ket{\alpha}\bra{\alpha}\bra{\lambda_{k}}  \e{i\, H_{int}T} \nonumber\\
	&= \sum_{j,k} c_j c_k^*\ket{\lambda_{j}}\ket{\alpha_j}\bra{\alpha_k}\bra{\lambda_{k}}. 
\end{align}

The reduction step is where the state changes non-unitarily to a state $\rho_i$ : 
\begin{align}
\rho(t) \rightarrow \rho_i=\frac{\mathbb{P}_i \rho(t) \mathbb{P}_i^\dagger}{Tr(\mathbb{P}_i\rho(t))}
\end{align}
Above, $ \mathbb{P}_i = |\lambda_i\rangle |a_i\rangle \langle a_i|\langle \lambda_i|$. 
Initially, in an ensemble of ${\cal SA}$s, each ${\cal SA}$ was in the same state. After the reduction, the ensemble becomes a mixed ensemble where the states are distributed within the ensemble according to the probability distribution $P(\lambda_i)=Tr(\mathbb{P}_i\rho(t))$ of getting outcome $\lambda_i$ on measurement of $O$. Then, if one picks an element randomly from the ensemble, the probability for getting a value $\lambda_{i}$ is $P(\lambda_{i})$. This process of the picking a system is the final step in a measurement.
If the information about which system was picked or equivalently the measured value of the observable is not accessible, then the state can be thought of as reducing to a statistical average $\rho_r= \sum_j P(\lambda_j) \rho_j= \sum_j \mathbb{P}_j \rho(t) \mathbb{P}_j^\dagger$. Note that this is not a fundamental feature of measurement and is essentially due to the inability of the observer to read out the values of measurement.

Probability distribution of an observable is completely specified by all it's moments. The moments of a local observable \footnote{By a local observable we mean an observable in the system Hilbert space} are the same when computed with the full density matrix or the reduced density matrix. Therefore, complete information about the probability distribution of a local observable of a system is contained in the reduced density matrix.  When one is interested in the local properties of the system, one can work with the reduced density matrix of the system, by tracing out the degrees of freedom of the apparatus
\begin{align}
\rho_{\op{S}}(T) &= \ptr{\op{A}}{\rho(T)} = \sum_n \bra{\alpha_{n}}\rho(T)\ket{\alpha_{n}}  \nonumber\\
&= \sum_{j} |c_j|^2 \ket{\lambda_{j}}\bra{\lambda_{j}}.
\end{align}

The discussion we had above can be generalized to the case of continuous eigenvalues, like position states. The initial state of the system $ \ket{\psi} = \int dx\, \psi(x) \ket{x} $ and the apparatus $ \ket{\alpha} $ evolve under the interaction Hamiltonian
\begin{align}
	H_{int} = \int dx\, \ket{x}\bra{x}\otimes A_x,
\end{align}
to the final density matrix
\begin{align}
	\rho(T) = \int dx\,d\xb\, \psi(x)\psi^{*}(\xb) \ket{x}\ket{\alpha_x}\bra{\alpha_{\xb}}\bra{\xb}.
\end{align}
We can trace out the apparatus, and obtain the system density matrix to be
\begin{align}
	\rho_{\op{S}}(T) = \int dx\, |\psi(x)|^2 \,\ket{x}\bra{x} = \int dx\,d\xb \, \psi(x)\psi^*(\xb)\delta(x-\xb) \ket{x}\bra{\xb}
\end{align}
One of the standout features of the reduced density matrix after sufficient time has elapsed is that it has become \textit{diagonal}, in other words the system is no longer in a superposition of states, and is referred to as \textit{decohered}. The entanglement and reduction process, where the environment was 'measuring' the position states of the system, leads to decoherence in that basis. The basis dependence of such processes has been part of intensive study. see for eg. \cite{schlosbook,TS,zurek1,zurekprd1}. As we are most interested in classicality and having position states naturally introduces the conjugate momentum states, it is worthwhile studying how the density matrix for this system behaves in the conjugate basis. To this end, we can change basis and obtain
\begin{align}
	\rho_{\op{S}}(T) &= \int dx\,d\xb\, dp\,d\pb\, \psi(x)\psi^*(\xb)\delta(x-\xb) \ket{p}\langle p|x\rangle\langle \xb|\pb\rangle\bra{\pb} \nonumber \\ 
	&= \int dp\,d\pb\, dx\, |\psi(x)|^2 \e{-i(p-\pb)x} \ket{p}\bra{\pb} \nonumber\\
	&= \int dp\,d\pb \, F[|\psi|^2; p,\pb] \ket{p} \bra{\pb},
\end{align}
where $ F[|\psi|^2; p,\pb] $ is the Fourier Transform of $ |\psi(x)|^2 $ and is a function of $ p $ and $ \pb $. The Fourier Transform in general receives support for all values of $ p $ and $ \pb $, i.e. not diagonal at all. What this shows is that the particular type of interaction that we were looking at, while it decohered the system in the $ x $-basis, maximally superposes it in the conjugate (momentum) basis. If we instead started with an interaction that measures the system's momentum, then it would decohere in the momentum basis while at the same time becoming maximally delocalized in the position basis.

\section{Smeared von Neumann measurements}
\label{s:von}
In the case of the interaction Hamiltonian we discussed above, the system decoheres in the basis associated with the interaction itself, while maximally superposing in the conjugate basis. Inherent in the form of the interaction Hamiltonian is the fact that the form of operator acting on the apparatus is determined exactly by the state of the system, i.e. if the system is in the state $ \ket{x} $, then the operator acting on the apparatus is $ A_{x} $. This would require an exact measurement of the system and that information used to infinitesimal detail on the apparatus, or in other words, the interaction needs correlations that are of infinitesimal precision. It is therefore natural to look at interaction between the system and apparatus that are motivated by the von Neumann form of the interaction, but take into account that we are using finite accuracy in the interaction by the use of a smearing function. To this end, we start by looking at the interaction Hamiltonian
\begin{align}
	H_{int} = \int dx\,dy\, g(x,y,\sigma) \, \ket{x}\bra{x} \otimes A_y,
\end{align}
where $ g(x,y,\sigma) $ is the smearing function that is based on a parameter $ \sigma $ that accounts for the finite accuracy. We set the convention that the $ \sigma\rightarrow 0 $ limit is the standard von Neumann interaction and $ g(x,y,0) = \delta(x-y) $. One another way to look at the same is that based on the state of the system, say $ \ket{x} $, the apparatus will undergo an evolution by $ A_y $ where $ y $ is determined by a function with mean at $ x $ and a standard deviation of $ \sigma $. Again, we will work with the assumption that $ H_{int} $ is the dominant term, and start with the density matrix $ \rho(0)  = \ket{\psi}\ket{\alpha} \bra{\alpha}\bra{\psi}$. Firstly, let us simplify the time evolution operator as
\begin{align}
	\e{-i H_{int} t} &= \exp\left( - i\,t\int dx\,dy\,g(x,y,\sigma)\,\ket{x}\bra{x}\otimes A_y \right) \nonumber\\
	&= \mathbf{1} -it\,\int dx\,dy\,g(x,y,\sigma)\,\ket{x}\bra{x}\otimes A_y  \nonumber\\ 
	&\ \ \  \ +\dfrac{(it)^2}{2!} \int dx_1\,dy_1\,dx_2\,dy_2\,g(x_1,y_1,\sigma)g(x_2,y_2,\sigma)\,\left[\ket{x_1}\bra{x_1}\otimes A_{y_1}\right] \left[\ket{x_2}\bra{x_2}\otimes A_{y_2}\right] + \dots \nonumber\\
	&= \int dx \, \left(\ket{x}\bra{x} \otimes \exp\left[-i\,t\,\int dy\, g(x,y,\sigma) \,A_y\right]\right) .
\end{align}
With this, we can time evolve $ \op{SA} $, and we obtain the density matrix
\begin{align}
	\rho(t) &= \int dx\,d\xb\, \psi(x)\psi^*(\xb) \ket{x}\ket{\alpha_{x,\sigma}(t)} \bra{\alpha_{\xb,\sigma}(t)}\bra{\xb}, \\
\text{where}\ \ \ \ket{\alpha_{x,\sigma}(t)} &= \exp\left[-i\,t\,\int dy\, g(x,y,\sigma) \,A_y\right] \ket{\alpha}.
\end{align}
Unlike in the earlier discussion where after sufficient time the apparatus states approached read-off states that were orthogonal, the smearing can be expected to take the apparatus to a state which is not necessarily an orthogonal state, but a superposition of such orthogonal states that is peaked around the $ x $-value with a spread of $ \sigma $. A nice convenient formulation for such a case can be expressed using the inner product as
\begin{align}
	\left\langle \alpha_{x,\sigma}(T)\,| \, \alpha_{\xb,\sigma}(T) \right\rangle = \dfrac{1}{\sqrt{2\pi\sigma^2}} \exp\left[-\frac{(x-\xb)^2}{2\sigma^2}\right].
\end{align}
For the system, we could compute the reduced density matrix computed after sufficiently long time (indicated by $ T $) would then give
\begin{align}
	\rho_{\op{S}}(T) &=  \ptr{A}{\int dx\,d\xb\, \psi(x)\psi^*(\xb) \ket{x}\ket{\alpha_{x,\sigma}(T)} \bra{\alpha_{\xb,\sigma}(T)}\bra{\xb}} \nonumber\\
	&= \lim\limits_{t\rightarrow\infty} \int dx\,d\xb\, \psi(x)\psi^*(\xb) \ket{x}\bra{\xb} \left\langle \alpha_{\xb,\sigma}(T)\,| \, \alpha_{x,\sigma}(T) \right\rangle \nonumber\\
	&= \dfrac{1}{\sqrt{2\pi\sigma^2}}\int dx\,d\xb\,	 \ee{-\frac{(x-\xb)^2}{2\sigma^2}}\psi(x)\,\psi^*(\xb) \ket{x}\bra{\xb}.
\end{align}
It can be easily checked that this has the right limit to the standard case, when $ \sigma\rightarrow 0 $. Crucially, we also will also need the expression for the reduced density matrix in the momentum basis, which can be obtained via a Fourier Transform as
\begin{align}
	\rho_{\op{S}}(T) &= \int \,\dfrac{dp\,d\pb}{2\pi}\, \left[\dfrac{1}{\sqrt{2\pi\sigma^2}}\int dx\,d\xb \ee{-\frac{(x-\xb)^2}{2\sigma^2}-ipx+ i\pb\xb}\psi(x)\,\psi^*(\xb) \right]\ket{p}\bra{\pb},
\end{align} 

where the term in the box-bracket is a function of $ p,\pb $ and dependent on the parameter $ \sigma $. We already have ensured that the $ \sigma\rightarrow 0 $ is identical to that of the standard von Neumann interaction, and the implications thereby would follow as in the previous section. The simplification of the term in the box-bracket by getting rid of the integrals is possible once the functional form of $ \psi(x) $ is defined. It is therefore illustrative to consider an prototypical wavefunction to understand the behaviour of the density matrices in the two different bases.

\subsection{Example: Gaussian Wavefunction}
\label{s:eg}
Let us define the Gaussian wavefunction as follows
\begin{align}
	\psi(x) = \dfrac{1}{(2\pi s^2)^{1/4}}\ee{-\dfrac{x^2}{4s^2}},
\end{align}
which will lead to the reduced density matrix in the position and momentum space, respectively, as
\begin{align}
	\rho_{\op{S}}(x,\xb,T) &= \dfrac{1}{\sqrt{4\pi^2 s^2 \sigma^2}} \ee{-\dfrac{(x-\xb)^2}{2\sigma^2}-\dfrac{(x^2+\xb^2)}{4s^2}}\\
	\rho_{\op{S}}(p,\pb,T) &= \dfrac{2 \sqrt{s^2}}{\sqrt{4s^2 +\sigma^2}} \ee{-\dfrac{(p-\pb)^2 (2 s^4)}{4s^2+\sigma^2} - \dfrac{ (p^2+\pb^2)s^2\sigma^2}{4s^2+\sigma^2}}
\end{align}

The reduced density matrices are a function of two parameters $ (s,\sigma) $, and it's effects are best studied by looking at the behaviour of the diagonal elements (namely $ x=\xb $ and $ p=\pb $), and the directions perpendicular to the diagonal elements (namely $ x=-\xb $ and $ p=-\pb $, which we will refer to as off-diagonal). 

Just looking at the functional behaviours, we have the following:
\begin{align}
	x = \xb \ \ \ \rho_s(x,x) &\sim \ee{-\frac{x^2}{2s^2}} \\
	x = -\xb \ \ \ \rho_s(x,-x) &\sim \ee{-x^2\big(\frac{2}{\sigma^2}+\frac{1}{2s^2}\Big)}\\
	p = \pb \ \ \ \rho_s(p,p) &\sim \ee{\frac{-p^2}{\big(\frac{2}{\sigma^2}+\frac{1}{2s^2}\Big)}}\\
	p = -\pb \ \ \ \rho_s(p,-p) &\sim \ee{-p^2 (2s^2)}
\end{align}

From these expressions, we can come up with the following observations from looking at certain limits of the parameters $ (s,\sigma) $:
\begin{itemize}

	\item $ \sigma \rightarrow 0,~ s\rightarrow \text{large}$: in the position basis, the  reduced density matrix $\rho_s$ gets localized to the diagonal, where it has a large spread. $\rho_s$ is almost diagonal in the momentum basis as well, along which it is spread very wide.
	\item $ \sigma\rightarrow 0, ~ s \rightarrow 0 $: in the position basis, $\rho_s$ is well localized along the diagonal and the off-diagonal directions; however,  in the momentum basis $\rho_s$ is very delocalized in all directions.
	\item $ \sigma \rightarrow\text{large}, ~ s \rightarrow 0  $: in the position basis, $\rho_s$ is very localized along the diagonal as well as off diagonal; in momentum basis it is spread out along both the diagonal and off diagonal.
	\item $ \sigma \rightarrow \text{large}, ~ s\rightarrow \text{large} $: in the position basis it is spread along the diagonal as well as off diagonal and localised in both in the momentum basis.
\end{itemize}

\begin{table}[ht]
\centering
 \begin{tabular}{|c|c|c|c|}
 \hline
 &&position&momentum\\
\hline
& diagonal & $\Leftarrow\Rightarrow$ &$\Leftarrow\Rightarrow$  \\
$ \sigma\rightarrow 0,s\rightarrow large$&off diagonal &$\Rightarrow\;\Leftarrow$&$\Rightarrow\;\Leftarrow$\\
 \hline
 & diagonal &$\Rightarrow\;\Leftarrow$  &$\Leftarrow\Rightarrow$\\
 $\sigma\rightarrow 0,s\rightarrow 0$&off diagonal &$ \Rightarrow\;\Leftarrow$ &$\Leftarrow\Rightarrow$\\
 \hline
  &diagonal &$\Rightarrow\;\Leftarrow$  &$\Leftarrow\Rightarrow$\\
 $\sigma\rightarrow large,s\rightarrow 0$&off diagonal &$\Rightarrow\;\Leftarrow$  &$\Leftarrow\Rightarrow$\\
 \hline
  &diagonal &$\Leftarrow\Rightarrow $ & $\Rightarrow\;\Leftarrow$ \\
 $\sigma\rightarrow large,s\rightarrow large$&off diagonal &$\Leftarrow\Rightarrow$  &$\Rightarrow\;\Leftarrow $\\
 \hline
 \end{tabular}
 \caption{Behavior of the reduced density matrix along the diagonal and the off diagonal direction for various values of $\sigma$ and  $s$. $\Leftarrow\Rightarrow$ indicates spread out and $\Rightarrow\;\Leftarrow$ denotes localized}
  \label{t:rho_beh}
 \end{table}

The above behavior is also pictorially represented in Table \ref{t:rho_beh}. It is clear from table \ref{t:rho_beh} that there is an inverse relation ship between the spread of $\rho_s$ along the diagonal in position basis and along the off diagonal in the momentum basis. This is true for the spread of $\rho_s$ along the off diagonal in position basis and along the diagonal in the momentum basis as well.
From the above example, it is evident that the reduced density matrix in the position and momentum basis can be brought into a form that looks \textit{almost} diagonal, with the particle having an \textit{almost}  well-defined postion and momentum. However, the term \textit{almost} needs  proper clarification, which we  shall give now. In the position space, since there exists an inherent inaccuracy in the measurement process which is parametrized by $\sigma$, $\rho_s$ makes sense only in a coarse grained setting. In other words, we can consider $\rho_s$ as a discretized object with a lattice spacing set by $\sigma$. It would therefore be irrelevant to talk about a precision within the context of the density matrix with length scales $\gtrsim \sigma$. In such a setting, if the wavefunction of the particle is such that $s >  \sigma$ , then it would have non-trivial support in multiple cells. However, considering $3s < \sigma$, say, $\sim 99.99\%$ of the probability will have support over a cell that would be centered around the mean, meaning that this system will be almost indinstinguishable to a wavefunction that was described by a delta function about the same mean \cite{bruckner}. Now, looking at the density matrix in the momentum basis, for $s = \sigma/N$, a similar localization idea can be talked about with the scale in the momentum space set by $ \hbar \sqrt{4+N^2} /\sigma)$, restoring the factors of $\hbar $. It is necessary to not let $N$ be very large, i.e. choose a wavepacket that is very localized, as it would lead to the binning in the momentum space to be large. To put some perspective with numbers, suppose $\sigma\sim 1\mu m$, with the particle localized well within that length scale, say $N=3$, then the momentum scale would be roughly $10^{-27}kgm/s$, which is the momentum of $10^6 $ protons provided that the velocity measurements are of the order of $1 \mu m/s$. With such coarseness in the measurements, it would not be possible to probe the microstructure of these particles anyway, and we can thus see how collective classicality emerges in an approximate sense with imprefect measurements.

\section{Summary and Outlook}

The standard von-Neumann measurements are particularly useful as a tool to describe environment induced decoherence and the measurement process itself. However, these also assume infinite accuracy in the measurement process, wherein the environment evolves to a read-off state that exactly describes the system. In a realistic setup this assumption can be relaxed, which is precisely the model that we have chosen. Along with it, we get the added benefit of analysis of measurement of conjugate quantities. Although we cannot get decoherence in an exact sense with these inaccurate measurements, it does model decoherence in an approximate sense; the conjugate quantities seem decohered for all intends and purposes. There is enough parametric freedom within these inaccurate measurements to argue that an emergent classicality of collection of systems is feasible.

In our analysis, we have worked under the simplification that the interaction term is the most dominant part of the Hamiltonian and neglect the system and environment components of the Hamiltonian. It would be interesting and worthwhile to systematically include these contributions to see if they would also result in describing emergent classical trajectories. It would also be interesting to find a construction of a probability density function which depend on both the conjugate variables using a Wigner transform or alike, such that the connections to classical dynamics could be understood further. We would like to return to these questions in a future work.

\textbf{Acknowledgements:}
We would like to thank A. P. Balachandran, Chethan Krishnan, Parveen kumar, Subramanya H for useful discussions. PNB and TPS gratefully acknowledge support from FRG scheme of National Institute of Technology Calicut.    

%
%


\begin{thebibliography}{99}
	\bibitem{schlosbook} M. Schlosshauer,  Decoherence
	and the Quantum-To-Classical Transition, Springer (2007); 
	\bibitem{schlos}M. Schlosshauer,  Decoherence, the measurement problem, and interpretations of quantum mechanics, Rev. Mod. Phys. 76, 1267 (2005), arxiv : quant-ph/0312059.
	\bibitem{TS} Tulsi Das, Measurements and Decoherence, arXiv:quant-ph/0505070 (2005)
	\bibitem{zeh}	H.D. Zeh, On the interpretation of measurement in quantum theory, Found. Phys. 1, 69 (1970).
	\bibitem{zurek physt} W.H.Zurek, Phys. Today 44(10), 36 (1990); revised version : ‘Deco-
	herence and the transition from quantum to classical-revisited’, Los
	Alamos Science, Number 27, 2002 (arxiv : quant-ph/0306072)
	
	\bibitem{zurekrmp2003} W.H. Zurek, ‘Decoherence, einselection and quantum origin of the classical’
	 Rev. Mod. Phys. 75, 715 (2003); (arxiv : quant-phy/0105127)
	
	\bibitem{joosbook} Joos et al (ed), ‘Decoherence and the appearance of a classical world
	in quantum theory’, second edition,  Springer (2003);
	\bibitem{zurek1} W. H. Zurek, Environment-induced superselection rules, Phys. Rev. D26, 1862–1880 (1982).
	\bibitem{zurekprl}Juan Pablo Paz  and Wojciech Hubert Zurek, Quantum Limit of Decoherence: Environment Induced Superselection of Energy Eigenstates, Phys. Rev. Lett, 82, 5181 (1999)
\bibitem{Bacc} G. Bacciagaluppi: ‘The role of decoherence in quantum theory’ in the
	Stanford Encyclopedia of Philisophy (Winter 2003 edition) ed. by E.N.
	Zalta; \url{http://plato.stanford.edu/arxives/win2003/entries/qm-decoherence/}
	\bibitem{zurekprd1}	W. H. Zurek, Pointer basis of quantum apparatus: Into what mixture does the wave packet collapse?, Phys. Rev. D 24, 1516 (1981)
	\bibitem{VNold} von Neumann, J., 1932, Mathematische Grundlagen der
	Quantenmechanik, (Springer, Berlin); reprinted 1981;
	English translation by R. T. Beyer, 1955: Mathematical
	Foundations of Quantum Mechanics, (Princeton University Press). 
	\bibitem{wheelerbook}	J.A. Wheeler and W.H. Zurek : ‘Quantum Theory and Measurement’,
	Princeton University Press (1983)	
	\bibitem{adler}	S.L. Adler: ‘Why decoherence has not solved the measurement problem:
	a response to P.W. Anderson’ arxiv: quant-ph/0112095
	\bibitem{barnett} Stephen M. Barnett, Quantum Information, Oxford University Press (2009)	
	\bibitem{peres} Asher Peres, Quantum Theory: Concepts and Methods, Kluwer Academic Publishers (2002)		
	\bibitem{bruckner} Johannes Kofler and Caslav Brukner, Classical World Arising out of Quantum Physics under the Restriction of Coarse-Grained Measurements, Phys. Rev. Lett. 99, 180403 (2007)
\end{thebibliography}
\end{document}